\documentclass[]{aa}
\usepackage{amsmath}
\usepackage{graphicx}
\usepackage{txfonts}

\begin{document}

   \title{What is wrong with steady accretion discs?}

   \author{C. J. Nixon
          \inst{1}
          \and
          J. E. Pringle\inst{1,2}
          }

   \institute{Department of Physics and Astronomy, University of Leicester, Leicester, LE1 7RH, UK\\
              \email{cjn@leicester.ac.uk}
         \and
             Institute of Astronomy, Madingley Road, Cambridge, CB3 0HA, UK
             }

   \date{\today}


   \abstract{In a standard, steady, thin accretion disc, the radial distribution of the dissipation of the accretion energy is determined simply by energy considerations. Here we draw attention to the fact that while the (quasi-)steady discs in dwarf novae in outburst are in agreement with the expected emission distribution, the steady discs in the nova-like variables are not. We note that essentially the only difference between these two sets of discs is the time for which they have been in the high viscosity, high accretion rate state. In such discs, the major process by which angular momentum is transported outwards is MHD turbulence. We speculate that such turbulence gives rise to corona-like structures (here called magnetically controlled zones, or MCZs) which are also able to provide non-negligible angular momentum transport,  the magnitude of which depends on the spatial scale $L$ of the magnetic field structures in such zones. For short-lived, high accretion rate discs (such as those in dwarf novae) we expect $L \sim H$ and the MCZ to have little effect. But, with time (such as in the nova-like variables) an inverse cascade in the MHD turbulence enables $L$, and the net effect of the MCZ, to grow. We present a simple toy model which demonstrates that such ideas can provide an explanation for the difference between the dwarf novae and the nova-like variable discs.}


   \maketitle

\section{Introduction}

In terms of the energy budget a steady accretion disc should be a relatively simple object. A dissipative process in the disc (the ``viscosity'') transfers angular momentum outwards, allowing the disc material to accrete. Simple energy and angular momentum considerations \citep[e.g.][]{Shakura:1973aa,Pringle:1981aa} imply that the rate of energy dissipation per unit area per unit time, $D(R)$, in a steady disc is given by
\begin{equation}
\label{standard}
D(R) = \frac{ 3 G M_\ast \dot{M}}{4 \pi R^3} \left[ 1 - \left( \frac{R_\ast}{R} \right)^{1/2} \right].
\end{equation}
Here $R$ is the (cylindrical) radius, $\dot{M}$ is the steady accretion rate, the central object has mass $M_\ast$ and radius $R_\ast$, and we have assumed that the rotation profile is Keplerian with angular frequency $\Omega = \sqrt{GM_\ast/R^3}$. We have also assumed a zero torque inner boundary condition at $R = R_\ast$, which in energetic terms is the equivalent of assuming that no rotational energy from the central object is fed into the disc through the boundary layer at the stellar surface \cite[cf.][]{Pringle:1977aa}. It is important to note that for a steady disc the quantity $D(R)$ does not depend on the (still uncertain) nature of the viscosity.

The accretion discs about which we have the most detailed information in terms of spatial detail and time-dependent behaviour are the discs in cataclysmic variables. These are short period semidetached binary systems in which a low mass quasi-main-sequence star fills its Roche lobe and transfers mass via an accretion disc to a companion white dwarf \citep[see, for example, the review by][]{Warner:1995aa}\footnote{In some cataclysmic variables the magnetic field of the white dwarf is strong enough to disrupt the disc far from the white dwarf surface, or prevent disc formation altogether. In this paper we focus on the non-magnetic cataclysmic variables where this effect is not present.}. In these systems the accretion discs most closely approximating steady discs are those to be found in the dwarf novae in outburst (and super-outburst) and in the nova-like variables\footnote{Note that none of these discs is exactly steady. What is required in the current context is that the time-scale for disc variation be much longer than the viscous time-scale for the inner disc regions where most of the energy is emitted. It is the spectrum of these inner disc regions that are of relevance here. Since the decline of dwarf novae from outburst is controlled by the viscous time-scale at the outer disc radii, this condition is well satisfied. Thus all of the discs described here are quasi-steady, in the sense that their inner regions should be described by the dissipation rate and distribution given in Equation 1.}. In these objects most of the emission in the optical and near ultraviolet (say, 600 -- 120 nm) comes from the accretion disc \citep[see, for example, the review by Bath \& Pringle, in][]{Pringle:1985aa}. In addition, the radiative properties of such discs should, in principle, be well understood, at least in their central regions from which most of the energy is emitted. These discs are fully ionized and have temperatures and densities at which emissivities and opacities are well known from stellar atmosphere computations. Even for the high measured values of disc viscosity ($\alpha \sim 0.2 - 0.3$ from fitting observed lightcurves with disc instability models; \citealt{Bath:1981aa,Pringle:1986aa,Smak:1999aa,Kotko:2012aa,Tetarenko:2018aa}; see also \citealt{King:2007aa,Martin:2019aa}) the transverse optical depths through the disc are typically high ($\tau \sim 10^2 - 10^3$; e.g. \citealt{Wade:1998aa}) so that provided the energy is dissipated close to the disc plane, where most of the mass, and therefore kinetic energy, is located \citep[as argued by, for example,][]{Shakura:1973aa} the emitted disc spectra should be well approximated in terms of radiating plane parallel atmospheres \citep{Wade:1998aa,Linnell:2007aa}.

In this paper we discuss the differences in the observed spectra of dwarf novae and the nova-like variables, and some of the suggested explanations for the spectral properties (Section~\ref{sec:spectra}). We then propose a toy model designed to explain the missing energy at short-wavelengths in the nova-like discs, and derive the resulting equations (Section~\ref{sec:newdisc}). In Section~\ref{sec:numerical} we provide numerical solutions of the equations and show the temperature profiles which result from this model. We also compute simple black-body spectra to show the effect on the energy budget. In Section~\ref{sec:discussion} we discuss our results in a wider context.

\section{Continuum Spectra}
\label{sec:spectra}

The continuum spectra of dwarf novae in outburst and of nova-like variables have been investigated by a number of authors. We concentrate here on the results relevant to the wavelength region, say, 120 -- 600 nm in which most of the disc energy is expected to be emitted. As mentioned above, of interest to us are the spectra of dwarf novae in outburst and of (non-magnetic) nova-like variables, and these should correspond to accretion discs, accreting at a steady rate.

\subsection{Dwarf Novae in Outburst}
In a statistical analysis of {\it IUE} spectra of dwarf novae and nova-like stars, \cite{la-Dous:1991aa} provides continuum spectra at {\it IUE} wavelengths, i.e. 125 -- 300 nm. For dwarf novae in outburst, the outburst spectra of those objects which are not close to edge on (\citealt{la-Dous:1991aa} looks at $i < 60^\circ$, where $i = 0^\circ$ is face-on) almost all lie more or less on top of each other when normalised at the arbitrary wavelength of 132 nm. She shows two well-observed objects (SS Cyg and VW Hyi) in her Fig. 3. She notes that the spectra are well described by the spectra of B2-3 V-III stars with $T_e \approx 20000 \pm 2000$\,K. This agrees with our expectation that the spectra of these objects is comfortably within the parameter range relevant to well-understood stellar spectra. \cite{la-Dous:1991aa} concludes that 80 per cent of the non-double eclipsing systems show essentially the same continuous flux distribution. She also shows that observations and theory show the same overall characteristics as far as the shape and slope of the continuum flux goes.

The most recent summary of the {\it IUE} archive is by \cite{Hamilton:2007aa}. They use the accretion disc spectra from the optically thick disc model grid of \cite{Wade:1998aa}. They find that in ``virtually every dwarf nova system in our {\it IUE} sample, the accretion disk by itself provided very good agreement with the outburst spectra.'' 

There are however some deviations in that in some systems the flux predicted by the disc model is lower than the observed flux longward of 170 nm. All these systems lie above the period gap, i.e. they are longer period systems with more massive and therefore more luminous secondaries.  They note that this discrepancy may be due to a contribution from the secondary star and/or the outermost disc boundary in their models being set at 10,000 K (in the longer period systems the disc is larger and so may have optically thick annuli at effective temperatures lower than $10^4$\,K).

In summary, the continuum spectra of dwarf novae in outburst are reasonably well fit by steady accretion disc atmosphere models, at least in the spectral region where most of the energy is expected to be emitted.

\subsection{Nova-like variables}
In contrast to this, \cite{la-Dous:1991aa} finds that although the nova-like objects are expected to display  a pattern roughly identical to that of the dwarf novae in outburst, they clearly do not. Almost all the continua in the nova-likes are decidedly redder than those in dwarf novae. This suggests that the observations contradict the view that the (non-magnetic) nova-like variables and the dwarf novae in outburst are just accretion discs in a stationary state.

We note that the conclusions of \cite{la-Dous:1991aa} confirmed earlier findings by \cite{Wade:1988aa}, also using {\it IUE} spectra, that nova-like systems systematically disagree with either black body or Kurucz (simulated stellar atmosphere) model accretion disc spectra, in that the model spectra were too blue.

Subsequent to {\it IUE}, there have been a number of UV-sensitive satellites launched, including the {\it Hopkins UV Telescope} ({\it HUT}), {\it FUSE} and {\it HST}. However, because of the limited flexibility of their observing schedules (in contrast to {\it IUE} where observing decisions could be made in real time -- see, for example, \citealt{Bath:1980aa}) all subsequent observations of ``steady accretion discs'' have been made only in nova-like variables. From the attempts at modelling the discs in these systems it is clear that a standard steady accretion disc is inadequate.

\cite{Long:1994aa} observe the bright nova-like variable IX Velorum with {\it HUT} in the range 83 -- 186 nm. They note that if the distance to IX Vel is $\sim 95$ pc, standard disc models without reddening cannot simultaneously reproduce the colour and flux in the UV. Improved fits to the data can be obtained by modifying the disc structure within three white dwarf radii. They get reasonable fits either by simply removing radiation from parts of the disc hotter than 42,000 K, by setting the disc temperature to be constant inside the radius where the temperature reaches 31,000 K, or by simply setting the inner radius of the disc to be at $R_{\rm min} \sim 2.6 R_{\rm WD} \sim 1.8 \times 10^9$ cm. \cite{Linnell:2007ab} undertake further modelling of this system using the spectral range 86 -- 170 nm and concur with \cite{Long:1994aa}. The authors state: ``It is possible to achieve a close synthetic spectrum fit to the observed {\it FUSE} and STIS spectra [with a mass transfer rate of ${\dot M} = 5 \times 10^{-9} M_\odot $/yr] by adopting an accretion disk temperature profile that agrees with the standard model except for the inner annuli, extending to $r/r_{\rm WD} \approx 4$. Shortward of that radius the accretion disk is approximately isothermal.''

\cite{Linnell:2007aa} also model {\it HST}, {\it FUSE} and SDSS spectra of SDSS J080908.39+381406.2. This object is a SW Sex type of cataclysmic variable. As described by \cite{Warner:1995aa} these are nova-like variables close to the top of the period gap, with periods of 3--4 hr. They are defined by strong single-peaked Balmer emission, and strong He II emission. They have high accretion rates.  From the modelling, \cite{Linnell:2007aa} conclude ``Extensive simulations demonstrate that the accretion disk must be truncated at its inner edge if the temperature profile follows the standard model''.

\cite{Linnell:2010aa} model the accretion disc in the nova-like variable RW Sextantis. Yet again the authors conclude that ``no standard model synthetic spectrum fits the observed spectrum.''

\cite{Godon:2017aa} analyse the UV spectra ({\it FUSE} \& {\it IUE}, 86--320 nm) of the three nova-like variables MV Lyr, BZ Cam and V592 Cas. They find that they can fit the spectra using the standard effective temperature formula, provided that they replace the inner radius $R_0 = R_{\rm WD}$ with $R_0 \approx 2 R_{\rm WD}$. They also point out that this implies that most of the accretion energy has gone missing, and give some suggestions as to where it might have gone (See Section~\ref{problem}). 

\subsection{The problem}
\label{problem}
There seems to be overwhelming evidence that while the steady accretion discs in dwarf novae in outburst can be modelled as standard discs, with the expected energy dissipation rate as a function of radius (Equation~\ref{standard}), the steady accretion discs in nova-like variables cannot. Attempts at modelling the spectra of these latter objects gives us a strong indication that the energy expected to be released from the innermost few radii is not observed. That is, in these objects, a sizeable fraction ($\sim 25$\%) of the disc accretion energy appears to be missing.

\cite{Balman:2014aa} and \cite{Godon:2017aa} suggest that in the nova-like variables at radii within a few $R_{\rm WD}$ the accretion flow becomes shocked to virial temperatures and predominantly non-radiative. It then manages to advect the accretion energy inwards and to deposit it below the white dwarf photosphere where it becomes thermalised. This suggestion has a number of drawbacks. \cite{Pringle:1979aa} argued that such flows could become shocked and virialised, but only at low accretion rates where the cooling timescales are sufficiently long. Such virialised gas has a temperature of $\sim 10^8$ K. If it is to settle onto the white dwarf it would need to cool, and would therefore release most of its energy at energies of 10 -- 20 keV. This is not observed.

\cite{Scepi:2018ac,Scepi:2018ab} find mass loss from the vertical boundaries of their shearing box simulations and conclude that this implies that magnetically driven winds might be important. Indeed they argue that models of cataclysmic variables need ``to be thoroughly revised to take into account that most of the accretion energy may be carried away by a wind instead of being locally dissipated''. Actually it has long been known that high inclination ($i \le 65^\circ$) cataclysmic variables with high accretion rates display emission lines with P Cygni profiles \citep{Cordova:1982aa,Drew:1987aa,Warner:1995aa}. The widths of the blue-shifted components reach velocities of 3,000--5,000 km/s which correspond to the escape velocity from close to the white dwarf, thus perhaps indicating flows from the inner disc or boundary layer. The general picture which emerges \citep{Knigge:1997aa,Proga:2003aa,Proga:2005aa,Matthews:2015aa} is that the wind is a moderately collimated, slowly accelerated outflow, is radiatively driven, and has a ratio of wind mass loss rate to disc accretion rate in the range $\dot{M}_{\rm wind}/\dot{M}_{\rm acc} \sim 0.001 - 0.1$. The implication of this work is that the winds do not carry away a large enough fraction of the accretion energy and thus do not have a significant effect on the energy budget of these objects.  Furthermore, there seems to be no difference between the winds observed in outbursting dwarf novae and in nova-like variables.\footnote{We might also add that \cite{Lubow:1994ab} show that even moderate angular momentum loss from a disc wind can lead to a disc instability, not seen in these objects.}

It is further evident that none of the proposed explanations for the absent flux in nova-like variables addresses the observation that the dwarf novae in outburst show no such problem. And yet, the discs in these two sets of objects should, in principle, be very similar. We suggest that this difference is the key to understanding this problem. We propose an explanation below (Section~\ref{sec:newdisc}).

\section{A modified disc model}
\label{sec:newdisc}

We have seen that in order to explain the properties of the steady accretion discs in both the outbursting dwarf novae and the nova-like variables some modification of standard accretion disc theory is required. The usual disc, outlined by for example \cite{Shakura:1973aa}, \cite{Pringle:1981aa}, has an internal kinematic viscosity $\nu$ which is responsible for all of the angular momentum transport, and all of the dissipation, which is therefore local.

We have noted that the continuum spectra from these discs should therefore be what one calculates by assuming a steady-state energy dissipation rate (Equation~1). The dwarf novae in outburst seem to agree with this quite well, whereas the nova-like spectra are noticeably redder. In order to fit the observed spectra with models, the observers conclude that it is necessary to modify the energy dissipation in the inner radii of the disc (roughly $R_{\rm WD} \le R \le 3 R_{\rm WD}$) by reducing the energy emitted from this region below what one expects from a steady-state disc. We therefore need to explain not only how such a reduction in observed emission is achieved, but also why it manifests itself in the nova-like variables but not in outbursting dwarf novae.

We begin by noting that all these discs are fully ionized and therefore that the viscosity is most likely due to fully developed MHD turbulence, powered by the magneto-rotational instability \citep[MRI;][]{Balbus:1991aa}. The turbulence is strong, with the dimensionless magnitude given by the Shakura-Sunyaev $\alpha$-parameter of order $\alpha \sim 0.2 - 0.3$ \citep{King:2007aa,Martin:2019aa}.

It has long been thought \citep[see, for example,][]{Galeev:1979aa} that by analogy with the solar surface, accretion discs might possess strongly magnetic layers at low optical depth. Indeed such hypothesized magnetic layers (there called ``coronae'') are standard in models to explain the high energy emission from accretion discs around black holes in X-ray binaries \citep[e.g.][]{White:1982aa} and active galactic nuclei \citep[AGN, e.g.][]{Haardt:1991aa}. We suggest here that such a strong magnetic layer might play a role here too. To describe the properties of such a layer, rather than use the terms such as ``corona'' or ``chromosphere'', which are loaded by use in solar, and other, contexts, we refer here to such a layer as a ``magnetically controlled zone'', henceforth MCZ.

We then also note that an obvious difference between the steady discs seen in outbursting dwarf novae, and those seen in the nova-like variables, is the length of time for which these discs have been in the high (fully ionized, high viscosity, large $\alpha$) state. The dwarf nova discs have been in the high state for a matter of days, whereas the nova-like variable discs have been in the high state for weeks to months. 

In addition, it has been suggested \citep[][see also \citealt{King:2004aa,Uzdensky:2008aa}]{Tout:1996aa} that while typical MHD disc turbulence gives rise to magnetic fields with typical spatial scales of order the disc thickness $\sim H$, as time progresses this structure can, through an inverse cascade process, give rise to magnetic fields with larger spatial scales, of order $\sim R$. Such processes have also been proposed for the generation of the large-scale fields present in the solar corona \citep{Hughes:2003aa}. Thus we propose here, that it is the typical length-scales of the magnetic fields in the MCZ that give rise to the differences between outbursting dwarf novae and the nova-like variables. In the dwarf novae, the discs have just entered the high state, and the MCZ fields have length scales $\sim H$, whereas the nova-like MCZs have had time (a few weeks) to develop fields in their MCZs which are larger, say $\sim R$.

In order to provide a tractable physical description of what the properties of such an MCZ might be, and how it might influence the properties of the underlying disc, it is necessary to make some simple assumptions and gross oversimplifications.\footnote{Given that the identification of the heating processes in the solar corona, and the description of its dynamics, is still an area of active research \citep[e.g.][]{Schrijver:2008aa,Hansteen:2015aa,Dahlburg:2016aa}, and that the MCZs of accretion discs are likely more extreme with dynamical shear and with rotational velocities comparable to the escape velocity, this seems a reasonable initial approach.} 

Thus we propose the following simple modification to standard disc theory. We shall assume that, unlike in a standard accretion disc, each disc annulus can transmit its angular momentum outwards in two ways:

\begin{enumerate}

\item In the usual manner, via a disc magneto-turbulent viscosity. This dissipates energy locally in the disc.

\item By using the poloidal field generated by the magneto-turbulence in the MCZ. We shall assume that the net effect of this process is  to launch material outwards to land further out in the disc. Such an idea was discussed by \cite{King:2007aa}. This process removes energy locally from the disc (radius $R$) and deposits it at some larger radius (say, $kR$, where $k \ge 1$).

\end{enumerate}

Thus we shall assume that when a fully ionized disc first forms, the dynamo-generated loops are small scale and thus we expect $k \approx 1 + O(H/R) \approx 1$. And as time progresses such a disc is able to generate larger scale poloidal fields, and thus $k$ increases until $k \approx 1 + O(1)$.

\subsection{The basic equations}
\label{sec:eqns}

With these ideas in mind, we now produce some modified disc evolution equations.
We shall do this in a manner which keeps assumptions and free parameters to a minimum. It will be evident that in order to do this we shall have to grossly simplify what we expect in reality to be going on. Nevertheless we hope that in this way we are able to capture the essence, and the net effects, of the physical processes that are involved.

We split the disc into a set of annuli of width $\Delta R$ \citep[cf.][]{Pringle:1981aa}. We assume that the effect of the magnetic fields is to enable mass to be lost from radius $R$ and to be deposited at a radius $kR$, where we assume $k \ge 1$. We assume that each parcel of mass $\delta m$ that is lost from radius $R$ takes with it an angular momentum (per unit mass) corresponding to $(kR)^2 \Omega(R)$ -- thus we assume that the magnetic field line along which the mass $\delta m$ flows is co-rotating with its footpoint at radius $R$ until it reaches a radius $kR$. We assume that the mass is then deposited at radius $kR$.

\subsubsection{Mass conservation}

We assume that the mass loss at radius $R$ per unit area is given by $\dot{\Sigma}_{\rm loss}(R)$, and that the corresponding mass gain at radius $R$, from an annulus at radius $R/k$, is given by $\dot{\Sigma}_{\rm gain}(R)$. Then the mass conservation equation becomes
\begin{equation}
\label{masscons}
\frac{\partial \Sigma}{\partial t} = - \frac{1}{R} \frac{\partial}{\partial R} (R \Sigma v_R)  - \dot{\Sigma}_{\rm loss}(R)  + \dot{\Sigma}_{\rm gain}(R) ,
\end{equation}
where $v_R$ is the radial mass flow velocity.

Recall that the mass being lost to the annulus, of width $\Delta R$ at radius $R$ is launched out to an annulus, of the same width $\Delta R$, at radius $kR$. Thus we have that
\begin{equation}
2 \pi (kR) \, \Delta R \, \dot{\Sigma}_{\rm gain}(kR) = 2 \pi R \, \Delta R\, \dot{\Sigma}_{\rm loss}(R).
\end{equation}

Thus, in equation(\ref{masscons}), we have that
\begin{equation}
\label{massgain}
\dot{\Sigma}_{\rm gain}(R) =  \left\{ \begin{array}{ll}
				k^{-1} \, \dot{\Sigma}_{\rm loss}(R/k) & \mbox{if $R > kR_{\rm WD}$ } \\
				0 & \mbox{otherwise}
			      \end{array}
			      \right. 
\end{equation}
Here we have allowed for the disc to terminate at the inner radius of $R_{\rm WD}$.\footnote{Note that in practice we use a logarithmic radial grid so that $\Delta R$ is a function of radius and must be carried through from equation (3) to equation (4). This is also true of equation (7) below.}

\subsubsection{Angular momentum conservation}
\label{sec:angmom}
Similarly \citep[cf.][]{Pringle:1981aa} we may write the angular momentum conservation equation as
\begin{eqnarray}
\label{angmom}
\frac{\partial}{\partial t}(\Sigma R^2 \Omega) & = & - \frac{1}{R} \frac{\partial}{\partial R} (R \Sigma v_R R^2 \Omega) - \dot{J}_{\rm loss}(R)\\
& & + \dot{J}_{\rm gain}(R) + \frac{1}{R} \frac{\partial}{\partial R}(\nu \Sigma R^3 \Omega^\prime). \nonumber
\end{eqnarray}
Our assumption, given above, implies that
\begin{equation}
\dot{J}_{\rm loss}(R) = \dot{\Sigma}_{\rm loss}(R) \, (kR)^2 \, \Omega(R),
\end{equation}
Thus, similarly, we have that
\begin{equation}
\dot{J}_{\rm gain}(R) =  \left\{ \begin{array}{ll}
				k^{-1} \, \dot{J}_{\rm loss}(R/k) \, R^2 \, \Omega(R/k) & \mbox{if $R > kR_{\rm WD}$ } \\
				0 & \mbox{otherwise}
			      \end{array}
			      \right. 
\end{equation}

Note that the mass leaving (arriving) at radius $R$ $(kR)$ does not carry the specific angular momentum of either radius. Therefore both annuli are no longer in centrifugal balance. Therefore additional mass must be shuffled inwards (outwards) to ensure centrifugal balance is restored and each ring is returned to Keplerian rotation \citep[see e.g.][for details]{Bath:1981aa,Bate:2010aa}.

\subsubsection{Energy dissipation}
The matter being deposited at radius $R$, originated at radius $R/k$. By assumption it reaches radius $R$ with the angular velocity corresponding to its original radius $R/k$. Thus it has acquired a circular velocity $R \Omega(R/k)$, and is being added to the local material which has circular velocity $R \Omega(R)$. 

Thus we may write the energy dissipation rate (per unit area per unit time) as
\begin{equation}
  \label{newenergy}
D(R) = \nu \Sigma (R \Omega^\prime)^2 + \dot{\Sigma}_{\rm gain}(R) \left( \frac{1}{2} \right) R^2 [\Omega(R/k) - \Omega(R)]^2,
\end{equation}
where the first term is the usual viscous dissipation and $\dot{\Sigma}_{\rm gain}(R)$ is given by Equation~(\ref{massgain}).

The shuffling process (Section~\ref{sec:angmom}), required to ensure centrifugal balance of each ring, also leads to energy dissipation both at radius $R$ and $R/k$. We find that the `shuffling energy' released at $R/k$ is comparable to, but generally smaller than the energy associated with equation (8). The `shuffling energy' released at radius $R$ is small. We include these additional heating terms into the energy balance equation (8).

\subsection{Dimensionless parameters}

We can define two dimensionless parameters which should determine the set up.

\begin{enumerate}

\item The parameter $k$ tells us how far the material is outwardly projected by the magnetic field. In the limit $k \rightarrow 1$ there should be no effect. This is presumably the same as assuming that $k-1 \approx H/R \ll 1$, that is that the scale-length of the magnetic field is $\approx H$. Once the magnetic field scale becomes of order $R$ we presumably have $k-1 \approx 1$.

\item The other parameter is concerned with the fraction of the mass flow that is involved in the MCZ. We can write this as the ratio between the mass loss rate at radius $R$, as being $\approx \pi R^2 \dot{\Sigma}_{\rm loss}(R)$, compared to the local disc accretion rate $\dot M$. Using the fact that in a steady disc (away from the boundaries) ${\dot M} = 3 \pi \nu \Sigma$, we may define the second parameter as
\begin{equation}
\label{massloss}
\mu = \frac{\dot{\Sigma}_{\rm loss}(R) R^2}{3 \nu \Sigma(R)}.
\end{equation}

\end{enumerate}

It is evident that $k$ will principally determine the radial scale of the effect, while $\mu$ will determine the magnitude of the effect. Below we explore varying these parameters and the effect they have on the disc structure.

\section{Numerical calculations}
\label{sec:numerical}
In this Section we present numerical calculations that solve the equations presented in Section~\ref{sec:eqns}. To do this we split the disc into concentric annuli, logarithmically spaced in radius so that $dR/R$ is a constant\footnote{This allows a constant $k$ to be associated with a constant integer $n_k$ which determines how many cells are between a radius $R$ and a radius $kR$.}. Typically we employ 250 annuli to model the disc. For the inner and outer boundaries we apply boundary conditions as follows. The inner boundary condition is the standard zero-torque condition which allows mass to flow freely off the inner boundary on to the central object, and for simplicity we do not add this mass to the central object. For the outer boundary we inhibit mass flow outwards, equivalent to applying a tidal torque from the companion star to the outer zone which keeps the mass within the disc.

To calculate the temperature of the disc at each radius we balance the local heating and cooling rates with the equation
\begin{equation}
  2\sigma_{\rm SB}T_{\rm eff}^4 = D(R) = \nu\Sigma(R\Omega^\prime)^2 + A(R)
\end{equation}
where $A(R)$ is the contribution to heating from material falling on to the disc from smaller radii due to the MCZ. The effective temperature is related to the midplane temperature as $T_{\rm eff}^4 = T_{\rm c}^4/\tau$ where $\tau = \kappa\Sigma/2$ is the optical depth and $\kappa$ is the opacity. For the opacity we employ Kramer's law with $\kappa = \kappa_0\rho T_{\rm c}^{-7/2}$ and $\kappa_0 = 5\times 10^{24}$ in cgs units \citep[see e.g.][]{Cox:1969aa,Cannizzo:1988aa,Frank:2002aa}.

For our model parameters, we take as a starting point the model for IX Vel given by \cite{Linnell:2007ab}. Their initial model takes $M_{\rm WD} = 0.8 M_\odot$, $R_{\rm WD} = 9.8459 \times 10^{-3} R_\odot$ and $\dot{M} = 5 \times 10^{-9} M_\odot$/yr. The runs of effective temperature they get from the unadjusted (poor fit, too much UV flux) and the adjusted models (good fit, with reduced inner disc temperatures) are given in their Table 7.

Then their final model produces the unadjusted and adjusted runs of effective temperature given in their Table 8. That has the same mass WD but a larger radius; $R_{\rm WD} = 0.015 R_\odot$. The details are given in Table 9, including assuming that the outer disc edge is at $R_{\rm out} = 0.33 D$, where the stellar separation is $D = 1.54 R_\odot$. We take the following parameters from their Table~9: $M_{\rm WD} = 0.8 M_\odot$, $r_{\rm WD} = 0.015 R_\odot$, ${\dot M} = 5.0\times 10^{-9}M_\odot/{\rm yr}$, $r_{\rm in} = 0.015 R_\odot$ and $r_{\rm out} = 0.56 R_\odot$.

\begin{figure*}
\begin{center}
\includegraphics[width=0.47\textwidth]{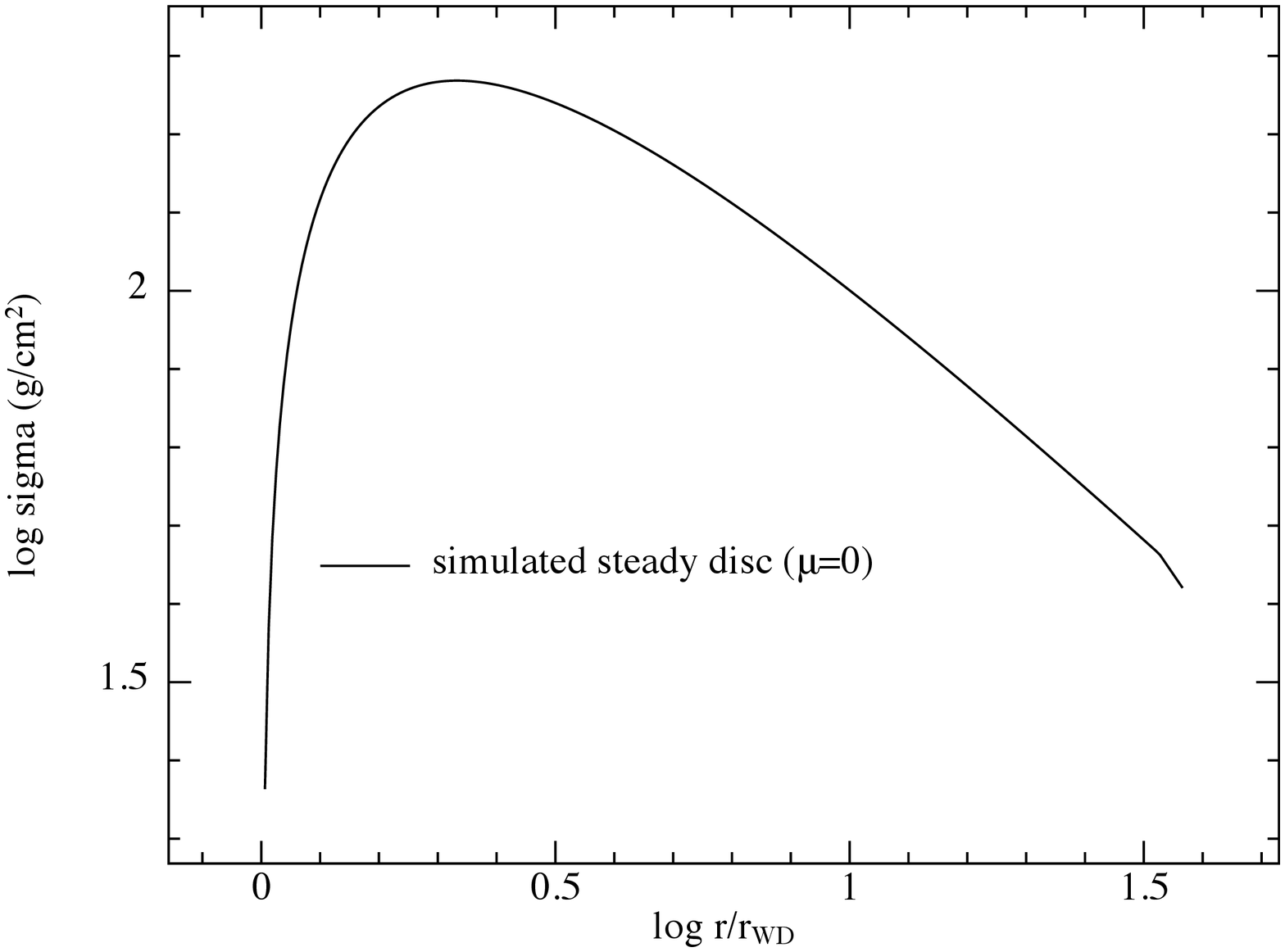}\hfill
\includegraphics[width=0.47\textwidth]{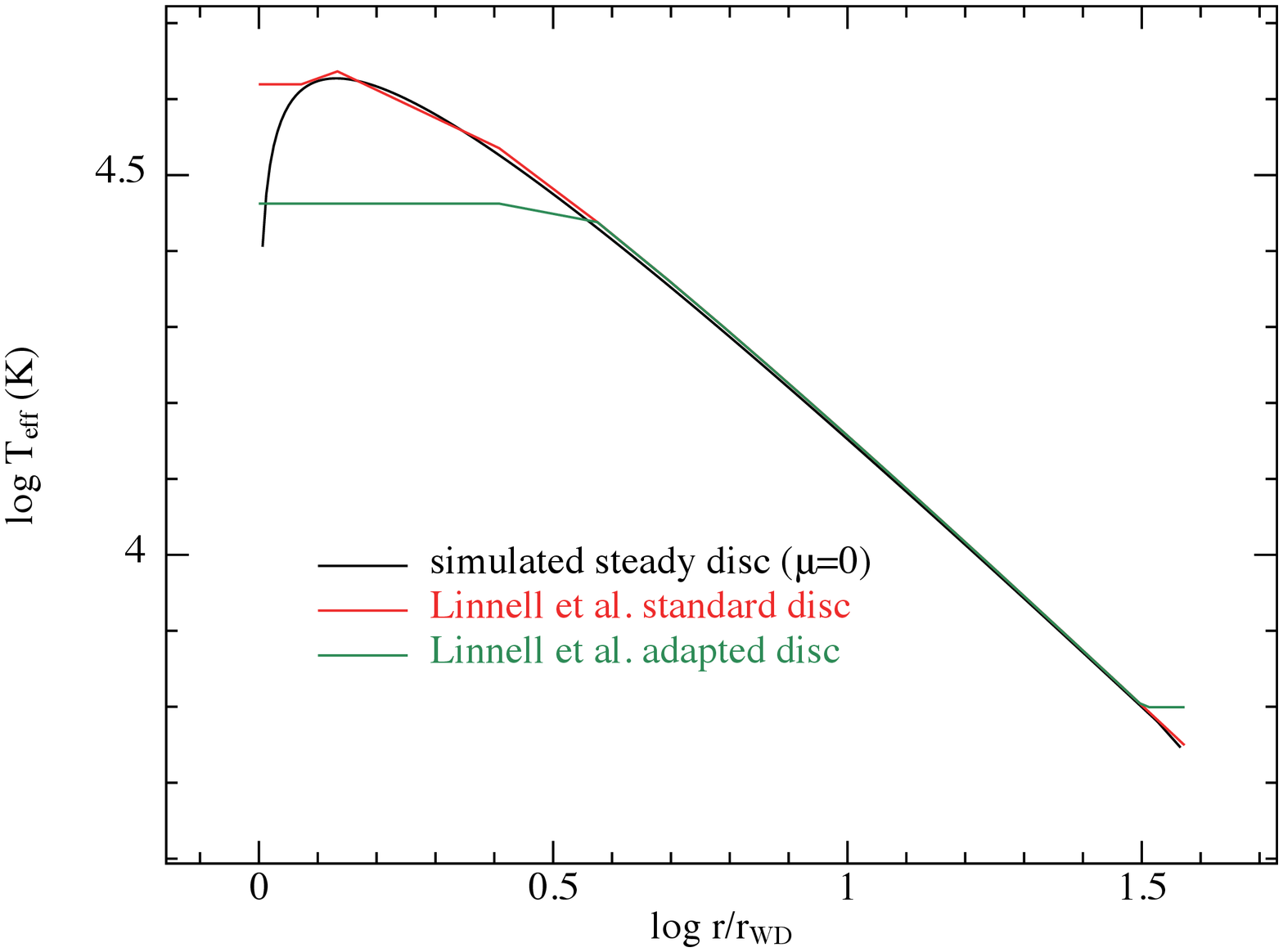}
\caption{These figures show the standard disc structure before the inclusion of the MCZ effects. Left hand panel: The surface density of the disc in g/cm$^2$ on a log scale plotted against the radius of the disc in units of the white dwarf radius. A small knee can be seen in the data at the location where mass is added to the disc at $0.9r_{\rm out}$. Otherwise the plot shows the expected power-law with an inner edge turnover. Right hand panel: The effective (surface) temperature profile of the disc in Kelvin. The peak temperature is approximately 42,400\,K. Overlaid in a red-solid line is the standard disc temperature profile taken from Table~8 of \cite{Linnell:2007ab}. Overlaid in a green line is the adjusted temperature model of \cite{Linnell:2007ab} which is isothermal inside a radius of $r/r_{\rm WD} = 2.56$.}\label{steady}
\end{center}
\end{figure*}

We start with the steady-state disc structure for the above parameters, and add mass to the disc with a rate ${\dot M} = 5.0\times 10^{-9}M_\odot/{\rm yr}$ at a radius of $0.9r_{\rm rout}$ smoothed with a cosine bell over a radial scale of $0.03r_{\rm out}$. We evolve this for a viscous timescale in the outer regions (approximately one month) to allow the disc to settle numerically (for this we set $k=1$ and $\mu=0$). During this phase we see no evolution of the disc, with the rate at which mass is added matched by the rate at which mass leaves the inner boundary of the disc (recall the outer boundary condition allows no mass to leave disc). The disc surface density and temperature profile are shown in Fig.~\ref{steady} and the data from Table~8 in \cite{Linnell:2007ab} are also plotted for comparison in the temperature plot.

\subsection{Results}
We start from the disc structure shown in Fig.~\ref{steady} and turn on the effects of the MCZ described by the equations in Section~\ref{sec:eqns}. We evolve the disc until a new steady-state is reached, which takes approximately a viscous timescale in the outer disc regions. We perform two sets of simulations to explore the effect of independently varying the two parameters $k$ and $\mu$.

\subsubsection{The effect of varying $k$}
First we fix $\mu = 0.1$ and vary $k$ between different simulations. To confirm the effect on the disc structure is small when the radial scale of the MCZ is small we run a simulation with $k=1.03$ which corresponds approximately to $k = 1 + H/R$. We then take values of $k = 1.25, 1.5, 2, 3$. We expect the scale and magnitude of the effect on the disc to increase with increasing $k$ and that $k \gtrsim 2$ will have a significant impact on the disc structure. \cite{Linnell:2007ab} require the temperature to be modified within $2-3$ white dwarf radii, so to explain this we are expecting to need $k \sim 2-3$. The results of these simulations are shown in Fig.~\ref{fixmu}. From this we can see that the MCZ has relatively little impact on the disc (for $\mu=0.1$) when $k\lesssim 1.25$. For $k=1.25$ the temperature profile retains its original shape but has decreased overall by approximately 10 per cent; similarly the surface density profile retains its original shape and is reduced by a factor of $\sim 1.5$. As $k$ is increased, we begin to see a noticeable deviation from the standard profile for $k\gtrsim 2$. A kink appears in both the surface density and temperature profile at the radial location $kR_{\rm WD}$. This is due to our assumption of a single value for the parameter $k$; annuli inside $kR_{\rm WD}$ lose mass to the MCZ but do not receive any mass, whereas annuli outside $kR_{\rm WD}$ both lose and gain mass from this effect. For $k=2$ there is a significant drop in the central temperature profile for annuli at $R < kR_{\rm WD}$, while outer radii ($R \gtrsim kR_{\rm WD}$) retain the standard temperature profile. The dissipation in these outer regions is now due to both viscous dissipation and dissipation from the material arriving due to the MCZ. For $k=3.0$ the energy dissipation from the MCZ is larger than than that due to viscous dissipation, resulting in the outer regions becoming hotter than the standard profile. For these cases with $k \gtrsim 2$, there is a significant radial velocity induced in the disc by the angular momentum transport driven by the MCZ, and this results in the surface density profile being significantly reduced. We comment on this further in the discussion below. These simulations show that while $k$ principally determines the scale of the effect of the MCZ, it also determines the magnitude of the effect with larger $k$ resulting in a progressively larger deviation from the standard model.

\subsubsection{The effect of varying $\mu$}
\begin{figure*}
\begin{center}
\includegraphics[width=0.47\textwidth]{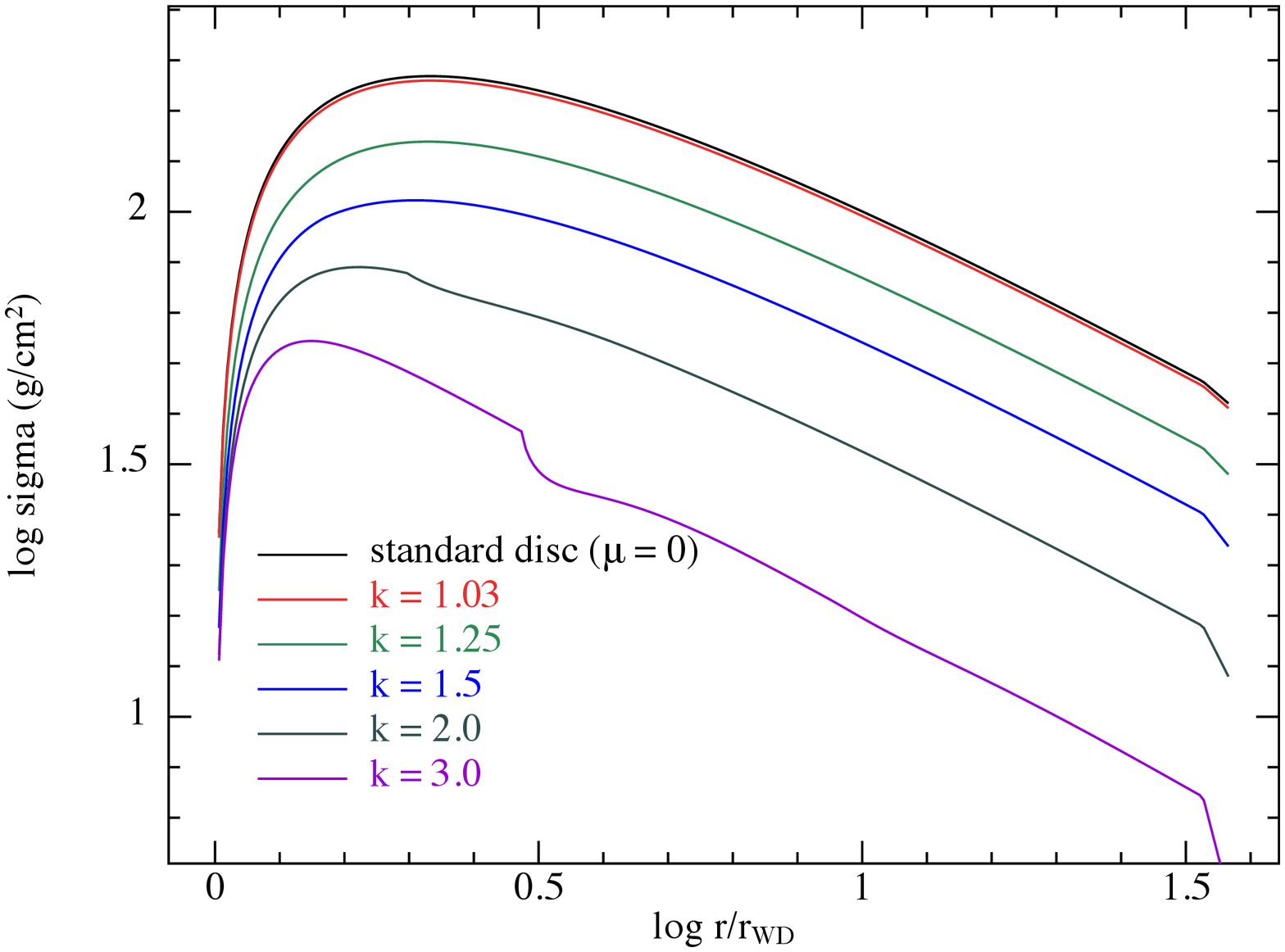}\hfill
\includegraphics[width=0.47\textwidth]{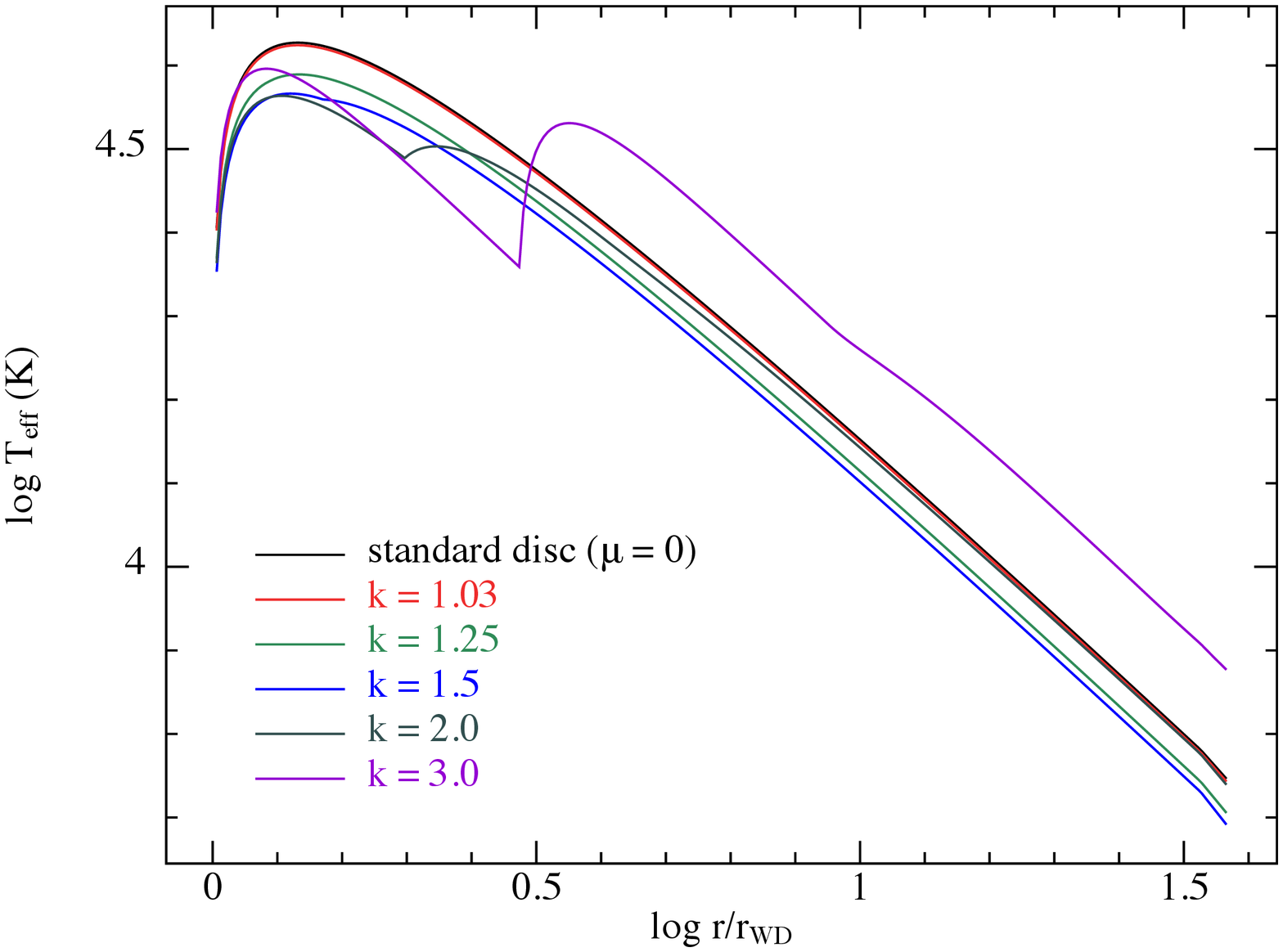}
\caption{MCZ models with a fixed vale of $\mu = 0.1$ and then $k$ varied as shown in the legend. The left hand panel shows the surface density, while the right hand panel shows the resulting temperature profiles.}\label{fixmu}
\end{center}
\end{figure*}

\begin{figure*}
\begin{center}
\includegraphics[width=0.47\textwidth]{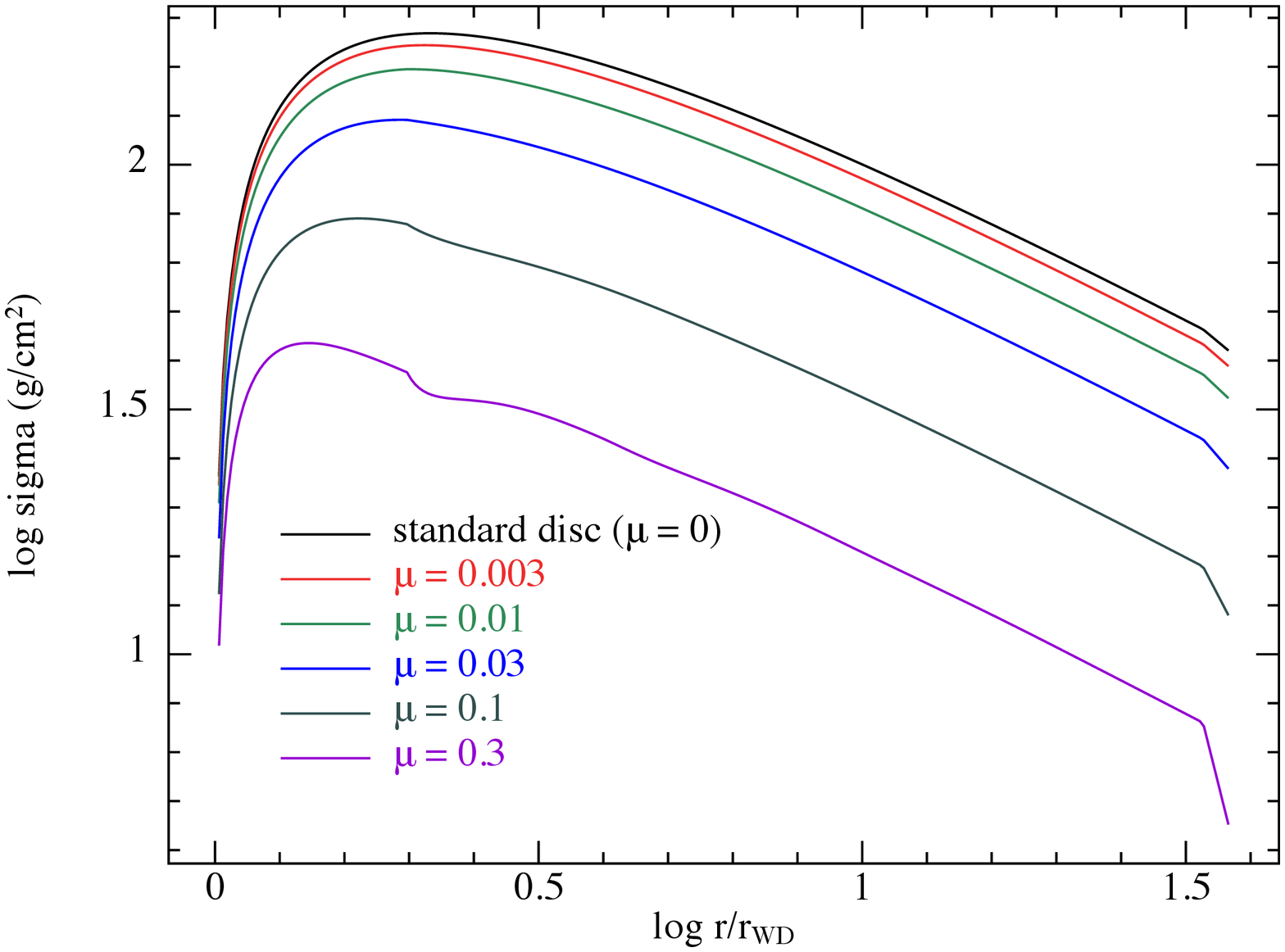}\hfill
\includegraphics[width=0.47\textwidth]{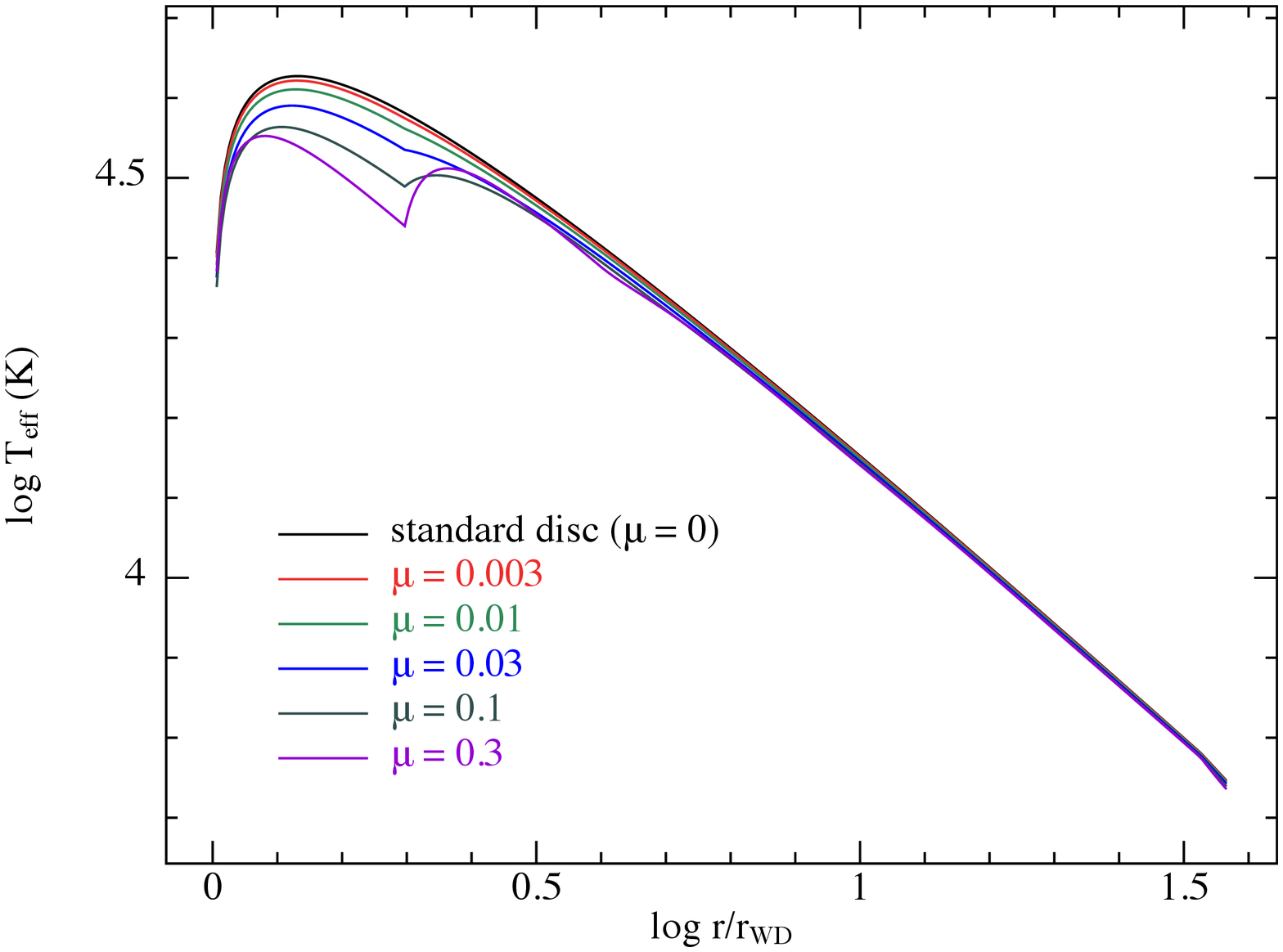}
\caption{Same as Fig.~\ref{fixmu} but this time holding $k=2$ fixed, and varying $\mu$. The left hand panel shows the surface density, while the right hand panel shows the resulting temperature profiles.}\label{fixk}
\end{center}
\end{figure*}

For the second set of simulations we fix $k$ and vary $\mu$. We take $k=2$ as this appears appropriate for generating the scale of effect required to fit the spectra of the nova-like variables. We then vary $\mu = 0.003,0.01,0.03,0.1,0.3$ finding that this range spans small variations to large variations from the standard profile. The results of these simulations are shown in Fig.~\ref{fixk}. In each case the radial scale of the effect remains the same, governed solely by the parameter $k$, while $\mu$ serves to scale the magnitude of the effect. For $\mu \lesssim 0.01$ both the surface density and temperature profiles are essentially unaltered. For $\mu = 0.1-0.3$ the surface density profile is globally lower than the standard case by a factor of $\approx 8-12$. This is also seen in Figure~\ref{fixmu}, and this implies a significantly enhanced radial velocity in the disc. For all of these cases the temperature profile of the disc at radii $R > kR_{\rm WD}$ is largely unaltered, but for $R < kR_{\rm WD}$ the disc effective temperature is reduced by an amount that increases with increasing $\mu$. For $\mu = 0.1-0.3$ the reduction in temperature in the inner regions is commensurate with that required by \cite{Linnell:2007ab}.

\subsubsection{What's required to fit the data?}
\begin{figure}
\begin{center}
\includegraphics[width=0.47\textwidth]{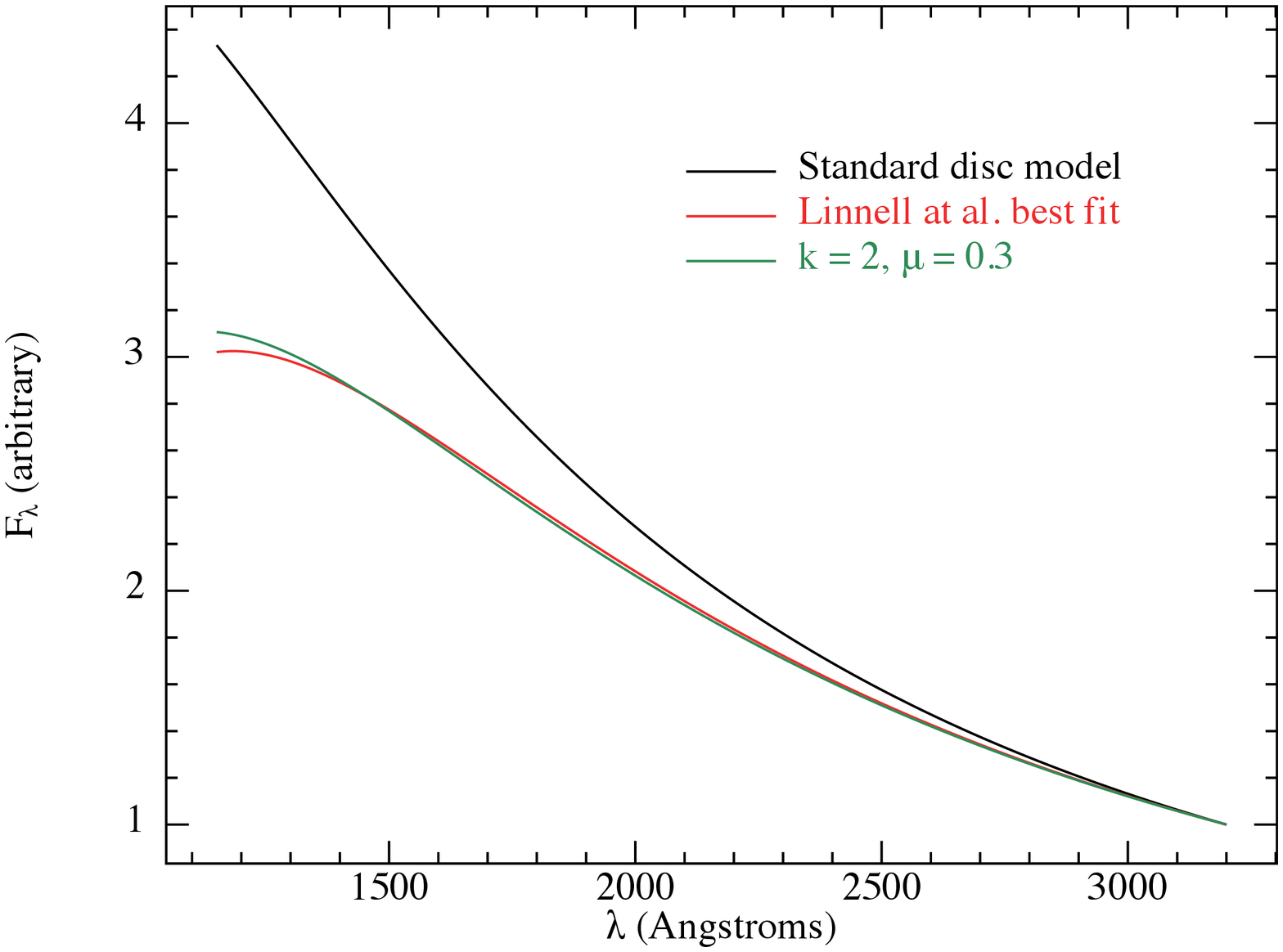}\hfill
\caption{Toy blackbody spectra constructed for the standard disc temperature profile (black), the \cite{Linnell:2007ab} best fit temperature profile (red) and the MCZ model presented here with $k = 2$ and $\mu = 0.3$ (green). The wavelength extent is chosen to match the data available for fitting and the wavebands in which the disc dominates the emission.}\label{specs}
\end{center}
\end{figure}

To achieve a significant effect on the temperature profile in the central regions we require a significant, although not unreasonable, amount of mass flow over a radial distance of order $R$. For a strong deviation of the temperature profile, of the magnitude required by \cite{Linnell:2007ab}, we need $\mu$ of order $0.1-0.3$ and $k \sim 2$. This is to be expected as the adapted model of \cite{Linnell:2007ab} leads to a drop in energy emitted by the disc of $\sim 25$ per cent. To reduce the energy dissipated in the inner disc regions by this amount (cf. equation~\ref{newenergy}) requires $\mu \approx 0.3$ for $k=2$. To confirm that these temperature profiles satisfy the energy budget requirements, we plot a simple blackbody spectrum for the $k=2,~\mu=0.3$ disc and also for the Linnell et al.\ adapted temperature profile. These are shown in Fig.~\ref{specs}. We note that the spectra in this region are significantly altered by line opacity effects, so a fit using stellar atmosphere grids to compute the spectra would be required to confirm a clear good fit to the data, but this is beyond our scope. However, Fig.~\ref{specs} confirms that we have succeeded in reducing the short-wavelength UV flux by the required amount and can produce a disc model which yields the right amount of energy as a function of wavelength to explain the spectra of nova-like variables.

\section{Discussion}
\label{sec:discussion}
We have proposed that the different emission patterns in the otherwise similar steady accretion discs in the outbursting dwarf novae and in the nova-like variables can be attributed to the active presence of magnetically controlled zones (MCZs), analogous to coronae/chromospheres in the solar and stellar contexts. These zones are located at low optical depths on either side of the disc plane. By the introduction of a simple ``toy model'' to mimic the possible affect of such zones on angular momentum transport, mass transport  and energy dissipation, we are able to provide a simple explanation of the difference between the two set of objects. In the dwarf novae, the length scales of the magnetic field in the MCZs are small (of order $\sim H$) and the MCZ has little effect, whereas at later times the MCZs in nova-like variables are able to develop larger scale fields $\sim R$ which are able to carry angular momentum (and energy) over larger distances. In these objects, the difference in emission patterns required by the observations can be accounted for by the MCZs in nova-like variables giving rise to a backflow of material (around $\mu \sim 10 - 30$ per cent of the overall accretion rate) on length-scales of order the local disc radius ($k \approx 2$). Observations require that the development of the larger scale fields in the MCZs take place on a timescale of weeks to months. We note here that such a change should be observable in systems with prolonged outbursts. \cite{la-Dous:1991aa} states that the sub-classes of SU UMa and Z Cam type dwarf novae, which exhibit outbursts that last weeks or months, cannot be distinguished from nova-like stars at UV wavelengths\footnote{We thank J.-P.~Lasota for drawing this to our attention.}. Another possibility is the VY Scl stars in the period after a long low state. Detailed analysis (similar to that of \citealt{Linnell:2007ab}) of these sub-classes at different epochs through the outbursts could provide strong constraints on the picture we present here.

There is some observational evidence for  the MCZ structures that we postulate here. First, it has long been known that non-negligible low optical depth emission is required in order to account for the fact that, in contrast to computed disc spectra, none of the objects display a Balmer Jump \citep{Wade:1984aa,Hassall:1985aa,la-Dous:1989aa}. Second, models of the emission line profiles in both outbursting dwarf novae and in nova-like variables seem to require not only a moderately collimated, supersonic wind-like outflow, but also a slower moving denser outward flow close to the disc photosphere \citep{Long:1996aa,Knigge:1997aa,Knigge:1997ab,Long:2002aa}. Third, there is evidence that in some cataclysmic variables and also in low-mass X-ray binaries (LMXBs) the disc thickness is larger than simple physics predicts \citep[see for example][]{Begelman:2007aa}.

It is worth noting that there are other contexts in which the standard steady accretion disc does not seem to readily account for the observations. For example, the observed time-delays (reverberation mapping) and microlensing in AGN seem to indicate that, as in the nova-like variables, the emission is coming from larger radii than standard disc theory would predict. This might be compatible with the properties of the discs with active MCZs that we describe here. In particular, compared to the usual steady accretion disc models, discs with active MCZs tend to emit the accretion energy at larger radii, and also to have lower surface densities (for a given accretion rate).

We also comment on the possible relationship between the properties of the MCZs postulated here, and those (disc coronae) proposed in the 1970s \citep{Shapiro:1976ab,Galeev:1979aa,Liang:1979aa} to account for the high energy emission observed in discs around black holes in low mass X-ray binaries and in AGN. It seems likely that the MCZs in all these objects are the outcomes of similar physical processes (as are, for example, the varied properties of stellar coronae in different types of stars). We note that there is as yet no complete explanation for the properties of, or the radial distribution of, such X-ray disc coronae. In all these objects, the underlying ability to generate an MCZ through the mechanism of fully developed MHD turbulence appears to be present. Thus whether or not a disc (or particular radial zones within a disc) is able to generate a substantial MCZ must depend on other factors. Possible physical effects at play here might include the effect of radiation pressure on disc stability \citep{Lightman:1974aa} and also the thermal (cooling) properties of the gas within the MCZ itself \citep[cf.][]{Pringle:1979aa,Mayer:2007aa,Rozanska:2015aa,Gronkiewicz:2019aa}.

In conclusion, it is evident that if the picture we describe in this paper has any relation to physical reality, then our understanding of the way in which accretion discs operate needs to be reassessed. For example, we can see from Figures 2 and 3 that as the importance of MCZs grows (increasing $k$ and/or $\mu$) the surface density, for a given net accretion rate, drops. This is because an additional means of transporting angular momentum implies an increased radial velocity. In this case, if the effects of an MCZ operate at large radii in the disc, then estimates of the turbulent viscosity parameter $\alpha$ that are based on the time-dependence of disc behaviour \citep[e.g.][]{Bath:1981aa,Kotko:2012aa} become unreliable. Similarly, it should be noted that the global behaviour of such MCZs imply that any attempt to try to estimate the nature and size of the viscosity $\alpha$ and the location and the nature of the dissipation of accretion energy by means of  numerical shearing-box techniques not only does not, but simply cannot, succeed. To investigate such phenomena it will be necessary to undertake global simulations of the kind that are already being carried out in numerical efforts to understand the dynamics of the solar corona \citep[see for example][]{Yeates:2007aa,Gibb:2014aa,Gibb:2016aa}.

\begin{acknowledgements}
We are grateful to the referee, Jean-Pierre Lasota, for a helpful report. CJN is supported by the Science and Technology Facilities Council (grant number ST/M005917/1). CJN thanks the Institute of Astronomy at the University of Cambridge for hospitality. We used {\sc splash} \citep{Price:2007aa} for the figures. This work was performed using the DiRAC Data Intensive service at Leicester, operated by the University of Leicester IT Services, which forms part of the STFC DiRAC HPC Facility (\url{www.dirac.ac.uk}). The equipment was funded by BEIS capital funding via STFC capital grants ST/K000373/1 and ST/R002363/1 and STFC DiRAC Operations grant ST/R001014/1. DiRAC is part of the National e-Infrastructure.
\end{acknowledgements}

\bibliographystyle{aasjournal}
\bibliography{nixon}

\begin{thebibliography}{}
\expandafter\ifx\csname natexlab\endcsname\relax\def\natexlab#1{#1}\fi
\providecommand{\url}[1]{\href{#1}{#1}}

\bibitem[{{Balbus} \& {Hawley}(1991)}]{Balbus:1991aa}
{Balbus}, S.~A., \& {Hawley}, J.~F. 1991, \apj, 376, 214

\bibitem[{{Balman} {et~al.}(2014){Balman}, {Godon}, \& {Sion}}]{Balman:2014aa}
{Balman}, {\c S}., {Godon}, P., \& {Sion}, E.~M. 2014, \apj, 794, 84

\bibitem[{{Bate} {et~al.}(2010){Bate}, {Lodato}, \& {Pringle}}]{Bate:2010aa}
{Bate}, M.~R., {Lodato}, G., \& {Pringle}, J.~E. 2010, \mnras, 401, 1505

\bibitem[{{Bath} \& {Pringle}(1981)}]{Bath:1981aa}
{Bath}, G.~T., \& {Pringle}, J.~E. 1981, \mnras, 194, 967

\bibitem[{{Bath} {et~al.}(1980){Bath}, {Pringle}, \& {Whelan}}]{Bath:1980aa}
{Bath}, G.~T., {Pringle}, J.~E., \& {Whelan}, J.~A.~J. 1980, \mnras, 190, 185

\bibitem[{{Begelman} \& {Pringle}(2007)}]{Begelman:2007aa}
{Begelman}, M.~C., \& {Pringle}, J.~E. 2007, \mnras, 375, 1070

\bibitem[{{Cannizzo} {et~al.}(1988){Cannizzo}, {Shafter}, \&
  {Wheeler}}]{Cannizzo:1988aa}
{Cannizzo}, J.~K., {Shafter}, A.~W., \& {Wheeler}, J.~C. 1988, \apj, 333, 227

\bibitem[{{Cordova} \& {Mason}(1982)}]{Cordova:1982aa}
{Cordova}, F.~A., \& {Mason}, K.~O. 1982, \apj, 260, 716

\bibitem[{{Cox} \& {Stewart}(1969)}]{Cox:1969aa}
{Cox}, A.~N., \& {Stewart}, J.~N. 1969, Nauchnye Informatsii, 15, 1

\bibitem[{{Dahlburg} {et~al.}(2016){Dahlburg}, {Einaudi}, {Taylor},
  {Ugarte-Urra}, {Warren}, {Rappazzo}, \& {Velli}}]{Dahlburg:2016aa}
{Dahlburg}, R.~B., {Einaudi}, G., {Taylor}, B.~D., {et~al.} 2016, \apj, 817, 47

\bibitem[{{Drew}(1987)}]{Drew:1987aa}
{Drew}, J.~E. 1987, \mnras, 224, 595

\bibitem[{{Frank} {et~al.}(2002){Frank}, {King}, \& {Raine}}]{Frank:2002aa}
{Frank}, J., {King}, A., \& {Raine}, D.~J. 2002, {Accretion Power in
  Astrophysics: Third Edition} (Cambridge University Press)

\bibitem[{{Galeev} {et~al.}(1979){Galeev}, {Rosner}, \&
  {Vaiana}}]{Galeev:1979aa}
{Galeev}, A.~A., {Rosner}, R., \& {Vaiana}, G.~S. 1979, \apj, 229, 318

\bibitem[{{Gibb} {et~al.}(2014){Gibb}, {Jardine}, \& {Mackay}}]{Gibb:2014aa}
{Gibb}, G.~P.~S., {Jardine}, M.~M., \& {Mackay}, D.~H. 2014, \mnras, 443, 3251

\bibitem[{{Gibb} {et~al.}(2016){Gibb}, {Mackay}, {Jardine}, \&
  {Yeates}}]{Gibb:2016aa}
{Gibb}, G.~P.~S., {Mackay}, D.~H., {Jardine}, M.~M., \& {Yeates}, A.~R. 2016,
  \mnras, 456, 3624

\bibitem[{{Godon} {et~al.}(2017){Godon}, {Sion}, {Balman}, \&
  {Blair}}]{Godon:2017aa}
{Godon}, P., {Sion}, E.~M., {Balman}, {\c S}., \& {Blair}, W.~P. 2017, \apj,
  846, 52

\bibitem[{{Gronkiewicz} \& {R{\'o}{\.z}a{\'n}ska}(2019)}]{Gronkiewicz:2019aa}
{Gronkiewicz}, D., \& {R{\'o}{\.z}a{\'n}ska}, A. 2019, arXiv e-prints,
  arXiv:1903.03641

\bibitem[{{Haardt} \& {Maraschi}(1991)}]{Haardt:1991aa}
{Haardt}, F., \& {Maraschi}, L. 1991, \apjl, 380, L51

\bibitem[{{Hamilton} {et~al.}(2007){Hamilton}, {Urban}, {Sion}, {Riedel},
  {Voyer}, {Marcy}, \& {Lakatos}}]{Hamilton:2007aa}
{Hamilton}, R.~T., {Urban}, J.~A., {Sion}, E.~M., {et~al.} 2007, \apj, 667,
  1139

\bibitem[{{Hansteen} {et~al.}(2015){Hansteen}, {Guerreiro}, {De Pontieu}, \&
  {Carlsson}}]{Hansteen:2015aa}
{Hansteen}, V., {Guerreiro}, N., {De Pontieu}, B., \& {Carlsson}, M. 2015,
  \apj, 811, 106

\bibitem[{{Hassall}(1985)}]{Hassall:1985aa}
{Hassall}, B.~J.~M. 1985, \mnras, 216, 335

\bibitem[{{Hughes} {et~al.}(2003){Hughes}, {Paczuski}, {Dendy}, {Helander}, \&
  {McClements}}]{Hughes:2003aa}
{Hughes}, D., {Paczuski}, M., {Dendy}, R.~O., {Helander}, P., \& {McClements},
  K.~G. 2003, Physical Review Letters, 90, 131101

\bibitem[{{King} {et~al.}(2007){King}, {Pringle}, \& {Livio}}]{King:2007aa}
{King}, A.~R., {Pringle}, J.~E., \& {Livio}, M. 2007, \mnras, 376, 1740

\bibitem[{{King} {et~al.}(2004){King}, {Pringle}, {West}, \&
  {Livio}}]{King:2004aa}
{King}, A.~R., {Pringle}, J.~E., {West}, R.~G., \& {Livio}, M. 2004, \mnras,
  348, 111

\bibitem[{{Knigge} \& {Drew}(1997)}]{Knigge:1997aa}
{Knigge}, C., \& {Drew}, J.~E. 1997, \apj, 486, 445

\bibitem[{{Knigge} {et~al.}(1997){Knigge}, {Long}, {Blair}, \&
  {Wade}}]{Knigge:1997ab}
{Knigge}, C., {Long}, K.~S., {Blair}, W.~P., \& {Wade}, R.~A. 1997, \apj, 476,
  291

\bibitem[{{Kotko} \& {Lasota}(2012)}]{Kotko:2012aa}
{Kotko}, I., \& {Lasota}, J.-P. 2012, \aap, 545, A115

\bibitem[{{la Dous}(1989)}]{la-Dous:1989aa}
{la Dous}, C. 1989, \mnras, 238, 935

\bibitem[{{la Dous}(1991)}]{la-Dous:1991aa}
---. 1991, \aap, 252, 100

\bibitem[{{Liang}(1979)}]{Liang:1979aa}
{Liang}, E.~P.~T. 1979, \apjl, 231, L111

\bibitem[{{Lightman} \& {Eardley}(1974)}]{Lightman:1974aa}
{Lightman}, A.~P., \& {Eardley}, D.~M. 1974, \apjl, 187, L1

\bibitem[{{Linnell} {et~al.}(2007{\natexlab{a}}){Linnell}, {Godon}, {Hubeny},
  {Sion}, \& {Szkody}}]{Linnell:2007ab}
{Linnell}, A.~P., {Godon}, P., {Hubeny}, I., {Sion}, E.~M., \& {Szkody}, P.
  2007{\natexlab{a}}, \apj, 662, 1204

\bibitem[{{Linnell} {et~al.}(2010){Linnell}, {Godon}, {Hubeny}, {Sion}, \&
  {Szkody}}]{Linnell:2010aa}
---. 2010, \apj, 719, 271

\bibitem[{{Linnell} {et~al.}(2007{\natexlab{b}}){Linnell}, {Hoard}, {Szkody},
  {Long}, {Hubeny}, {G{\"a}nsicke}, \& {Sion}}]{Linnell:2007aa}
{Linnell}, A.~P., {Hoard}, D.~W., {Szkody}, P., {et~al.} 2007{\natexlab{b}},
  \apj, 654, 1036

\bibitem[{{Long} \& {Knigge}(2002)}]{Long:2002aa}
{Long}, K.~S., \& {Knigge}, C. 2002, \apj, 579, 725

\bibitem[{{Long} {et~al.}(1996){Long}, {Mauche}, {Raymond}, {Szkody}, \&
  {Mattei}}]{Long:1996aa}
{Long}, K.~S., {Mauche}, C.~W., {Raymond}, J.~C., {Szkody}, P., \& {Mattei},
  J.~A. 1996, \apj, 469, 841

\bibitem[{{Long} {et~al.}(1994){Long}, {Wade}, {Blair}, {Davidsen}, \&
  {Hubeny}}]{Long:1994aa}
{Long}, K.~S., {Wade}, R.~A., {Blair}, W.~P., {Davidsen}, A.~F., \& {Hubeny},
  I. 1994, \apj, 426, 704

\bibitem[{{Lubow} {et~al.}(1994){Lubow}, {Papaloizou}, \&
  {Pringle}}]{Lubow:1994ab}
{Lubow}, S.~H., {Papaloizou}, J.~C.~B., \& {Pringle}, J.~E. 1994, \mnras, 268,
  1010

\bibitem[{{Martin} {et~al.}(2019){Martin}, {Nixon}, {Pringle}, \&
  {Livio}}]{Martin:2019aa}
{Martin}, R.~G., {Nixon}, C.~J., {Pringle}, J.~E., \& {Livio}, M. 2019, \na,
  70, 7

\bibitem[{{Matthews} {et~al.}(2015){Matthews}, {Knigge}, {Long}, {Sim}, \&
  {Higginbottom}}]{Matthews:2015aa}
{Matthews}, J.~H., {Knigge}, C., {Long}, K.~S., {Sim}, S.~A., \&
  {Higginbottom}, N. 2015, \mnras, 450, 3331

\bibitem[{{Mayer} \& {Pringle}(2007)}]{Mayer:2007aa}
{Mayer}, M., \& {Pringle}, J.~E. 2007, \mnras, 376, 435

\bibitem[{{Price}(2007)}]{Price:2007aa}
{Price}, D.~J. 2007, \pasa, 24, 159

\bibitem[{{Pringle}(1977)}]{Pringle:1977aa}
{Pringle}, J.~E. 1977, \mnras, 178, 195

\bibitem[{{Pringle}(1981)}]{Pringle:1981aa}
---. 1981, \araa, 19, 137

\bibitem[{{Pringle} \& {Savonije}(1979)}]{Pringle:1979aa}
{Pringle}, J.~E., \& {Savonije}, G.~J. 1979, \mnras, 187, 777

\bibitem[{{Pringle} {et~al.}(1986){Pringle}, {Verbunt}, \&
  {Wade}}]{Pringle:1986aa}
{Pringle}, J.~E., {Verbunt}, F., \& {Wade}, R.~A. 1986, \mnras, 221, 169

\bibitem[{{Pringle} \& {Wade}(1985)}]{Pringle:1985aa}
{Pringle}, J.~E., \& {Wade}, R.~A. 1985, {Interacting binary stars} (Cambridge
  University Press)

\bibitem[{{Proga}(2003)}]{Proga:2003aa}
{Proga}, D. 2003, \apjl, 592, L9

\bibitem[{{Proga}(2005)}]{Proga:2005aa}
{Proga}, D. 2005, in Astronomical Society of the Pacific Conference Series,
  Vol. 330, The Astrophysics of Cataclysmic Variables and Related Objects, ed.
  J.-M. {Hameury} \& J.-P. {Lasota}, 103

\bibitem[{{R{\'o}{\.z}a{\'n}ska} {et~al.}(2015){R{\'o}{\.z}a{\'n}ska},
  {Malzac}, {Belmont}, {Czerny}, \& {Petrucci}}]{Rozanska:2015aa}
{R{\'o}{\.z}a{\'n}ska}, A., {Malzac}, J., {Belmont}, R., {Czerny}, B., \&
  {Petrucci}, P.-O. 2015, \aap, 580, A77

\bibitem[{{Scepi} {et~al.}(2018{\natexlab{a}}){Scepi}, {Dubus}, \&
  {Lesur}}]{Scepi:2018ac}
{Scepi}, N., {Dubus}, G., \& {Lesur}, G. 2018{\natexlab{a}}, arXiv e-prints,
  arXiv:1812.02076

\bibitem[{{Scepi} {et~al.}(2018{\natexlab{b}}){Scepi}, {Lesur}, {Dubus}, \&
  {Flock}}]{Scepi:2018ab}
{Scepi}, N., {Lesur}, G., {Dubus}, G., \& {Flock}, M. 2018{\natexlab{b}}, \aap,
  620, A49

\bibitem[{{Schrijver} \& {Zwaan}(2008)}]{Schrijver:2008aa}
{Schrijver}, C.~J., \& {Zwaan}, C. 2008, {Solar and Stellar Magnetic Activity}
  (Cambridge University Press)

\bibitem[{{Shakura} \& {Sunyaev}(1973)}]{Shakura:1973aa}
{Shakura}, N.~I., \& {Sunyaev}, R.~A. 1973, \aap, 24, 337

\bibitem[{{Shapiro} {et~al.}(1976){Shapiro}, {Lightman}, \&
  {Eardley}}]{Shapiro:1976ab}
{Shapiro}, S.~L., {Lightman}, A.~P., \& {Eardley}, D.~M. 1976, \apj, 204, 187

\bibitem[{{Smak}(1999)}]{Smak:1999aa}
{Smak}, J. 1999, \actaa, 49, 391

\bibitem[{{Tetarenko} {et~al.}(2018){Tetarenko}, {Lasota}, {Heinke}, {Dubus},
  \& {Sivakoff}}]{Tetarenko:2018aa}
{Tetarenko}, B.~E., {Lasota}, J.~P., {Heinke}, C.~O., {Dubus}, G., \&
  {Sivakoff}, G.~R. 2018, \nat, 554, 69

\bibitem[{{Tout} \& {Pringle}(1996)}]{Tout:1996aa}
{Tout}, C.~A., \& {Pringle}, J.~E. 1996, \mnras, 281, 219

\bibitem[{{Uzdensky} \& {Goodman}(2008)}]{Uzdensky:2008aa}
{Uzdensky}, D.~A., \& {Goodman}, J. 2008, \apj, 682, 608

\bibitem[{{Wade}(1984)}]{Wade:1984aa}
{Wade}, R.~A. 1984, \mnras, 208, 381

\bibitem[{{Wade}(1988)}]{Wade:1988aa}
---. 1988, \apj, 335, 394

\bibitem[{{Wade} \& {Hubeny}(1998)}]{Wade:1998aa}
{Wade}, R.~A., \& {Hubeny}, I. 1998, \apj, 509, 350

\bibitem[{{Warner}(1995)}]{Warner:1995aa}
{Warner}, B. 1995, {Cataclysmic variable stars} (Cambridge University Press)

\bibitem[{{White} \& {Holt}(1982)}]{White:1982aa}
{White}, N.~E., \& {Holt}, S.~S. 1982, \apj, 257, 318

\bibitem[{{Yeates} {et~al.}(2007){Yeates}, {Mackay}, \& {van
  Ballegooijen}}]{Yeates:2007aa}
{Yeates}, A.~R., {Mackay}, D.~H., \& {van Ballegooijen}, A.~A. 2007, \solphys,
  245, 87

\end{thebibliography}




\end{document}